\documentclass[useAMS,usenatbib]{mn2e}
\usepackage{graphicx,amssymb}
\newcommand{\vc}[1]{\bmath{#1}} 
\newcommand{\lx}[1]{\rmn{#1}}

\newcommand{\figref}[2]{figure~\ref{#1}{\em #2}}
\newcommand{\secref}[1]{section~\ref{#1}}

\newcommand{\apref}[1]{Appendix~\ref{#1}}
\newcommand{\Eqref}[1]{Equation~(\ref{#1})}
\newcommand{\eqref}[1]{equation~(\ref{#1})}
\newcommand{\eqsref}[1]{equations~(\ref{#1})}
\newcommand{\Eqsand}[2]{Equations~(\ref{#1}) and (\ref{#2})}
\newcommand{\eqsand}[2]{equations~(\ref{#1}) and (\ref{#2})}

\newcommand{\eqsdash}[2]{equations~(\ref{#1}--\ref{#2})}
\newcommand{\exref}[1]{(\ref{#1})}
\newcommand{\bea}{\begin{eqnarray}}
\newcommand{\eea}{\end{eqnarray}}
\newcommand{\beq}{\begin{equation}}
\newcommand{\eeq}{\end{equation}}
\newcommand{\lt}{\left}
\newcommand{\rt}{\right}
\newcommand{\dd}{\partial}
\newcommand{\did}{\lx{d}}

\newcommand{\eps}{\epsilon}

\newcommand{\mfp}{\lambda_{\lx{mfp}}} 

\newcommand{\vth}{v_{\lx{th}}}

\renewcommand{\Re}{\lx{Re}}
\newcommand{\pperp}{p_\perp}
\newcommand{\pperpi}{p_{\perp i}}
\newcommand{\pperpe}{p_{\perp e}}
\newcommand{\dpperp}{\delta\pperp}

\newcommand{\dpperpe}{\delta\pperpe}
\newcommand{\ppar}{p_\parallel}
\newcommand{\ppari}{p_{\parallel i}}
\newcommand{\ppare}{p_{\parallel e}}
\newcommand{\dppar}{\delta\ppar}

\newcommand{\dppare}{\delta\ppare}
\newcommand{\vdel}{\vc{\nabla}}
\newcommand{\vdperp}{\vdel_\perp}

\newcommand{\vv}{\vc{v}}
\newcommand{\vperp}{v_\perp}
\newcommand{\vvperp}{\vv_\perp}
\newcommand{\vpar}{v_\parallel}
\newcommand{\vr}{\vc{r}}
\newcommand{\vk}{\vc{k}}
\newcommand{\vkperp}{\vk_\perp}
\newcommand{\kperp}{k_\perp}
\newcommand{\kpar}{k_\parallel}
\newcommand{\vu}{\vc{u}}
\newcommand{\vueff}{\vu_{\rm eff}}
\newcommand{\dvu}{\delta\vu}
\newcommand{\dvuperp}{\dvu_\perp}
\newcommand{\dupar}{\delta u_\parallel}
\newcommand{\va}{\vc{a}}
\newcommand{\vE}{\vc{E}}
\newcommand{\vEg}{\vc{\tilde E}}

\newcommand{\vj}{\vc{j}}
\newcommand{\vB}{\vc{B}}
\newcommand{\dvB}{\delta\vc{B}}
\newcommand{\dvBperp}{\dvB_\perp}
\newcommand{\dB}{\delta B}
\newcommand{\dn}{\delta n}

\newcommand{\dBpar}{\delta B_\parallel}
\newcommand{\vb}{\vc{b}}
\newcommand{\dvb}{\delta\vb}
\newcommand{\gmax}{\gamma_{\rm max}}
\newcommand{\St}[1]{C[#1]}
\newcommand{\vT}{\textbfss{T}}
\newcommand{\vP}{\textbfss{P}}
\newcommand{\vS}{\textbfss{S}}
\newcommand{\dV}{\delta\vc{V}}
\newcommand{\vG}{\textbfss{G}}
\newcommand{\vsig}{\vc{\sigma}}
\newcommand{\vI}{\textbfss{I}}

\newcommand{\qperp}{q_\perp}
\newcommand{\qperpi}{q_{\perp i}}

\newcommand{\qpar}{q_\parallel}
\newcommand{\qpari}{q_{\parallel i}}

\newcommand{\dqperp}{\delta\qperp}
\newcommand{\dqpar}{\delta\qpar}
\newcommand{\vq}{\vc{q}}
\newcommand{\vqperp}{\vq_\perp}
\newcommand{\vqpar}{\vq_\parallel}
\newcommand{\gT}{\Gamma_T}
\newcommand{\DD}{\delta}
\newcommand{\obar}{{\bar\omega}}
\newcommand{\kp}{k_\lx{p}}

\title[Firehose and gyrothermal instabilities]{Magnetofluid dynamics of magnetized cosmic plasma: firehose and gyrothermal instabilities}
\author[A. A. Schekochihin et al.]{
  A. A. Schekochihin,$\!^{1}$\thanks{E-mail: a.schekochihin1@physics.ox.ac.uk}
  S. C. Cowley,$\!^{2,3}$
  F. Rincon,$\!^{4}$
  and M. S. Rosin$^{5}$\\
  $^{1}$ Rudolf Peierls Centre for Theoretical Physics, University of Oxford, 
  1 Keble Road, Oxford, OX1 3NP, U.K.\\
  $^{2}$ EURATOM/CCFE Association, Culham Science Centre,
  Abingdon, OX14 3DB, U.K.\\
  $^{3}$ Blackett Laboratory, Imperial College, 
  Prince Consort Road, London, SW7 2AZ, U.K.\\
  $^{4}$ Laboratoire d'Astrophysique de Toulouse-Tarbes, Universit\'e de Toulouse, CNRS, 
  14 avenue Edouard Belin, F-31400 Toulouse, France\\
  $^{5}$ DAMTP, Centre for Mathematical Sciences, University of Cambridge, 
  Wilberforce Road, Cambridge, CB3 0WA, U.K.
}
\begin{document}

\date{E-print {\tt arXiv:0912.1359}; published in {\it MNRAS} {\bf 405}, 291 (2010)}

\pagerange{\pageref{firstpage}--\pageref{lastpage}} \pubyear{2009}

\maketitle

\label{firstpage}

\begin{abstract}
Both global dynamics and turbulence in magnetized weakly collisional 
cosmic plasmas are described by general magnetofluid equations 
that contain pressure anisotropies and heat fluxes that must be 
calculated from microscopic plasma kinetic theory. It is shown that 
even without a detailed calculation of the pressure anisotropy or the heat fluxes, 
one finds the macroscale dynamics to be generically unstable to microscale 
Alfv\'enically polarized fluctuations. Two instabilities that can be treated 
this way are considered in detail: the parallel firehose instability (including 
the finite-Larmor-radius effects that determine the growth rate 
and scale of the fastest growing mode) and the {\em gyrothermal instability (GTI)}. 
The latter is a new result --- it is shown that a parallel ion heat flux 
destabilizes Alfv\'enically polarized fluctuations even in the 
absence of the negative pressure anisotropy required for the firehose.
The main physical conclusion is that both pressure anisotropies 
and heat fluxes associated with the macroscale dynamics trigger 
plasma microinstabilities and, therefore, their values 
will likely be set by the nonlinear evolution of these instabilities. 
Ideas for understanding this nonlinear evolution are discussed. 
It is argued that cosmic plasmas will generically be ``three-scale systems,'' 
comprising global dynamics, mesoscale turbulence and microscale 
plasma fluctuations. The astrophysical example of cool cores of galaxy 
clusters is considered quantitatively and it is noted that 
observations point to turbulence in clusters (velocity, 
magnetic and temperature fluctuations) being in 
a marginal state with respect to plasma microinstabilities and so 
it is the plasma microphysics that is likely to set the heating 
and conduction properties of the intracluster medium. 
In particular, a lower bound on the scale of temperature fluctuations 
implied by the GTI is derived. 
\end{abstract}


\section{Introduction}

Many astrophysical plasmas are not sufficiently collisional 
to be described by the standard fluid equations of magnetohydrodynamics (MHD)
\citep[see, e.g.,][]{Balbus_MVI,Schek06,Sharma06,Sharma07}. 
When the collision frequency $\nu$ is smaller than the Larmor frequency 
$\Omega=eB/mc$ of the particle gyration about the magnetic-field lines, 
the plasma becomes {\em magnetized}: pressure and heat flux 
are now tensors that depend on the local direction 
of the magnetic field. This complication leads to three significant 
physical effects. Firstly, on the macroscopic scales, 
the momentum and heat transport become highly anisotropic 
with respect to the magnetic-field direction. 
Secondly, old MHD instabilities, like the MRI, that are believed to excite 
turbulence in astrophysical systems \citep{Balbus_review}, 
are significantly modified \citep{Quataert_kinMRI,Sharma_kinMRI,Islam} 
and new ones appear: MTI \citep{Balbus_MTI}, MVI \citep{Balbus_MVI}, 
HBI \citep{Quataert_HBI}. 
Thirdly, a host of superfast {\em microscale} instabilities exist
that are directly driven by the pressure anisotropies 
\citep[see][and references therein]{Schek05,Sharma06}
and, as we are about to discover, also by heat fluxes.  

The presence of microscale instabilities especially opens a fundamental 
problem: the equations one tends to use to describe the macroscopic dynamics 
of magnetized plasma, be they fluid or kinetic, are derived 
in the long-wavelength limit \citep[$k\rho\ll1$, where 
$\rho$ is the Larmor radius; see][]{Kulsrud} and turn out to be ill-posed 
because in this limit the microinstabilities have growth rates proportional 
to $k$ \citep{Schek05}. 
In order to regularize them at small scales, one has to take 
into account effects associated with the finite Larmor radius (FLR), which 
requires fairly complicated kinetic theory and typically means 
that the full multiscale problem is analytically hard and 
numerically intractable. Ideally, one would like to have an effective 
mean-field theory, with the microscale fluctuations 
analytically averaged to produce some form of closure for the momentum 
and heat transport. This has not been achieved yet, but 
an educated guess about the form of such a closure 
can me made, based on the idea that the system should always find itself 
in the marginal state with respect to the microinstabilities 
\citep{Sharma06,Sharma07,Schek06,Lyutikov,Kunz}. 

In this paper, we attempt to make progress in setting up the theoretical 
framework for astrophysical plasma dynamics by addressing three basic questions: 
what is the general form of the dynamical equations that 
we are attempting to approximate? what can be learned about the 
microinstabilities under the most general assumptions? 
what constraints do their marginal stability conditions impose on the 
allowed macroscopic states of the plasma? 
The first of these questions is addressed in \secref{sec:eqns}, 
the second in \secref{sec:lin}, where an old (firehose) and a new 
(gyrothermal) instabilities of Alfv\'enically polarized perturbations 
are derived. Possible ways of thinking 
about the nonlinear physics of these microinstabilities are proposed 
in \secref{sec:nlin}. The physical conclusions are summarized 
in \secref{sec:disc}, including a discussion of 
the relevance of all this in galaxy cluster cores (as a case study 
of a multiscale astrophysical plasma system). 

\section{Equations for plasma dynamics}
\label{sec:eqns}

Let us consider a two-species fully ionized plasma. 
In the completely general case (assuming only quasineutrality), 
the evolution of ion density $n$ and flow velocity $\vu$ is governed 
by the following equations 
\bea
\label{eq:cont}
\lefteqn{\frac{\did n}{\did t} = -n\vdel\cdot\vu,}&&\\
\lefteqn{mn\,\frac{\did\vu}{\did t} 
= -\vdel\cdot\lt(\vP + \vI\,\frac{B^2}{8\pi} - \frac{\vB\vB}{4\pi}\rt),}&& 
\label{eq:mom}
\eea
where $m$ is the ion mass, $\did/\did t = \dd/\dd t + \vu\cdot\vdel$ 
the convective derivative, $\vI$ the unit dyadic, $\vB$ the magnetic 
field and $\vP$ the plasma pressure tensor. It is via $\vP$ that 
all the kinetic physics comes in: in general, $\vP$ is the sum 
of the ion and electron pressures and for each species, it is 
$\vP = m\int d^3\vv\,\vv\vv f$, calculated 
from the distribution function $f(t,\vr,\vv)$, which is the solution 
of the kinetic equation for that species. Note that $\vv$ is the 
peculiar velocity, i.e., the particle's velocity in a frame moving with 
the mean flow velocity $\vu$.

Thus, the challenge is to calculate $\vP$. 
This typically involves setting up an asymptotic expansion of the kinetic 
equation with respect 
to one or several of the small parameters available for the plasma 
under the set of macroscopic conditions of interest.
Many such expansions for magnetized plasma exist, corresponding to various 
physical regimes: collisional \citep{Braginskii,Mikh1,Mikh2,Catto}, 
long-wavelength collisionless, or drift-kinetic \citep{CGL,Kulsrud}, 
short-wavelength anisotropic, or gyrokinetic 
\citep[][and references therein]{Howes,Tome}, 
and more specialized versions of the above, appropriate for the 
treatment of pressure-anisotropy-driven instabilities: 
firehose \citep{Schek08,Rosin} and mirror \citep{Califano,Istomin,Rincon}. 
We do not at the moment wish to pick any one of these, 
but simply notice that in all of them, the equilibrium distribution 
function invariably turns out to be gyrotropic, i.e., 
independent of the phase angle of the particle's Larmor gyration. 
The only assumptions needed for that is that the characteristic 
frequencies $\omega$ 
for the evolution both of the equilibrium and of the perturbations thereof 
should be smaller than the ion Larmor frequency $\Omega$ 
and the length scales of the equilibrium longer than the ion 
Larmor radius $\rho$. If the pressure tensor is assumed 
to be determined purely by the gyrotropic lowest-order distribution, 
then it reduces to a diagonal form, 
$\vP = \pperp(\vI-\vb\vb) + \ppar\vb\vb$, where $\vb=\vB/B$, 
and the perpendicular and parallel pressures are 
$\pperp = m\int\did^3\vv\,(\vperp^2/2)f$ and
$\ppar = m\int\did^3\vv\,\vpar^2f$.
These pressures can be shown to satisfy the so-called 
CGL equations: for each particle species, they are 
(\citealt{CGL,Kulsrud,Snyder1}; see \apref{app:KMHD} for a
simple derivation)
\bea
\label{eq:pperp}
\lefteqn{\pperp\frac{\did}{\did t}\ln\frac{\pperp}{nB}=
-\vdel\cdot\vqperp - \qperp\vdel\cdot\vb
-\nu(\pperp-\ppar),}&&\\
\label{eq:ppar}
\lefteqn{\ppar\frac{\did}{\did t}\ln\frac{\ppar B^2}{n^3}= 
-\vdel\cdot\vqpar + 2\qperp\vdel\cdot\vb
-2\nu(\ppar-\pperp),}&&
\eea
where $\nu$ is the collision frequency, 
$\qperp = m\int\did^3\vv\,\vpar(\vperp^2/2)f$ 
and $\qpar = m\int\did^3\vv\,\vpar^3f$ are the parallel 
fluxes of the perpendicular and parallel heat 
and $\vqperp=\vb\qperp$, $\vqpar=\vb\qpar$. 

As mentioned above, pressure anisotropies 
$\pperp-\ppar\neq0$ lead to instabilities whose peak 
growth rates occur at scales smaller than those 
allowed by the validity of the diagonal 
approximation for $\vP$ and are not captured by this 
approximation \citep{Schek05}. The instabilities are regularized 
by the FLR effects, so it is natural to resort to FLR 
corrections in the plasma pressure tensor \citep{Snyder2,Ramos,Passot}.
To lowest order in $\omega/\Omega$ and $k\rho_i$, this is
quite easy to do and the result, a simple derivation 
of which is given in \apref{app:P}, is
\bea
\label{eq:Pfull}
\vP &=& \pperp\vI - (\pperp-\ppar)\vb\vb + \vG,\\
\nonumber
\vG &=& \frac{1}{4\Omega}\bigl[\vb\times\vS\cdot\lt(\vI + 3\vb\vb\rt)
- \lt(\vI + 3\vb\vb\rt)\cdot\vS\times\vb\bigr]\\
&&+\,\frac{1}{\Omega}\bigl[\vb\lt(\vsig\times\vb\rt) + \lt(\vsig\times\vb\rt)\vb\bigr],
\label{eq:G}
\eea
where the auxiliary tensor $\vS$ and vector $\vsig$ are
\bea
\label{eq:S_def}
\vS &=& \lt(\pperp\vdel\vu + \vdel\vqperp\rt) + \lt(\pperp\vdel\vu + \vdel\vqperp\rt)^T,\\
\vsig &=& (\pperp-\ppar)\lt(\frac{\did\vb}{\did t} + \vb\cdot\vdel\vu\rt) 
+ (3\qperp-\qpar)\vb\cdot\vdel\vb.
\label{eq:sig_def}
\eea
Each plasma species contributes a pressure tensor of the form \exref{eq:Pfull}. 
In general, electron pressures are comparable to ion pressures, but it is not hard 
to show that the electrons' contribution to the FLR term $\vG$ is smaller 
than the ions' by a factor of $(m_e/m_i)^{1/2}$. 

Note that if one sets $\pperp-\ppar=0$ and $3\qperp-\qpar=0$ (as would be the case 
for an isotropic equilibrium distribution and collisional heat fluxes), 
the FLR term $\vG$ in \eqref{eq:Pfull} is readily recognized as the so called 
``gyroviscosity'' tensor, first obtained (in the collisional limit) 
by \citet{Braginskii} (he assumed sonic flows and found just 
the $\vdel\vu$ terms; the heat flux terms were introduced later 
by \citet{Mikh1,Mikh2} to accommodate subsonic flows). 

Thus, the momentum equation \exref{eq:mom} has the form
\bea
\nonumber
mn\,\frac{\did\vu}{\did t} 
&=& -\vdel\lt(\pperp + \frac{B^2}{8\pi}\rt)\\ 
&&+\, \vdel\cdot\lt[\vb\vb\lt(\pperp-\ppar + \frac{B^2}{4\pi}\rt) - \vG\rt],
\label{eq:vu}
\eea
where $\vG$ is given by \eqref{eq:G}. 
Now we need an evolution equation for the magnetic field.
Faraday's law reads
\beq
\label{eq:Faraday}
\frac{\dd\vB}{\dd t} = -c\vdel\times\vE,
\eeq
where $\vE$ is the electric field. The electron momentum equation 
is used to calculated $\vE$. Since the electron mass is small 
compared to the ion mass, to lowest order in $(m_e/m_i)^{1/2}$ 
this reduces to the force balance
\beq
\label{eq:fbal}
-\vdel\cdot\vP_e - en_e\lt(\vE + \frac{\vu_e\times\vB}{c}\rt) = 0. 
\eeq
The electron density $n_e$ is related to the ion density $n$ by 
the quasineutrality of the plasma, $n_e=Zn$ (the ion charge is $Z$ times 
electron charge $e$). 
The electron flow velocity $\vu_e$ is related to the ion flow velocity 
$\vu$ by $\vu_e=\vu -\vj/en_e$, where, using Amp\`ere's law, 
the current density is $\vj=c\vdel\times\vB/4\pi$. 
Finally, since the FLR terms in the electron pressure 
tensor are negligible to lowest order in $(m_e/m_i)^{1/2}$, 
we have $\vP_e = \pperpe\vI - (\pperpe-\ppare)\vb\vb$.
Assembling all this together, we get\footnote{See \apref{app:flux} 
for the demonstration that this equation conserves magnetic flux 
except at very small scales, where electron pressure anisotropy 
can lead to violation of flux freezing.}
\bea
\label{eq:ind}
\frac{\did\vB}{\did t} &=& \vB\cdot\vdel\vu - \vB\vdel\cdot\vu
-c\vdel\times\vEg,\\
\nonumber
\vEg &=& -\frac{1}{en_e}\vdel\lt(\pperpe + \frac{B^2}{8\pi}\rt)\\
&& + \frac{1}{en_e}\vdel\cdot\lt[\vb\vb\lt(\pperpe-\ppare + \frac{B^2}{4\pi}\rt)\rt].
\eea
Note that $c/en_e = B/mn\Omega$, where $m$, $n$ and $\Omega$ are ion 
mass, density and Larmor frequency, respectively. 

We will not be preoccupied here with the determination of the 
pressures and heat fluxes
(which is necessary to close the set of equations we have written down). 
Depending on the physical regime one is interested in, 
they can either be calculated in the collisional 
limit \citep{Braginskii} or Landau fluid closures can 
be devised for them, appropriate for a collisionless plasma 
\citep{Snyder1,Snyder2,Ramos,Passot}. 
Instead of wading into this rather complex subject, 
we will inquire what can be learned just from the 
general form of the equations of plasma dynamics outlined above.

\section{Firehose and gyrothermal instabilities}
\label{sec:lin}

In any given astrophysical problem, one might find some macroscale  
solution of the equations of \secref{sec:eqns}, describing the 
large-scale dynamics. Such solutions turn out to be 
generically unstable to perturbations with large wavenumbers 
and high frequencies (much larger than the fluid turnover rates 
$\omega\gg|\vdel\vu|$). In general, showing this involves having 
to perturb all quantities, including the pressures and the heat 
fluxes, which requires a kinetic closure. However, there is a class 
of perturbations whose stability does not depend on the details 
of kinetic theory. 

Let us start by perturbing the momentum equation \exref{eq:vu}. 
We assume the perturbation to be $\propto \exp (-i\omega t + i\vk\cdot\vr)$. 
In our perturbation theory, we will always consider terms containing 
$\omega$ and $\vk$ to be dominant in comparison with the terms 
containing time derivatives or gradients of the macroscale quantities. 
Thus, from \eqref{eq:vu}, we get, noting that $\vdel\cdot\vB=0$ implies 
$\vdel\cdot\vb = -\vb\cdot\vdel B/B$, 
\bea
\nonumber
mn\omega\dvu &=& \vkperp\lt(\dpperp + \frac{B\dB}{4\pi}\rt) + 
\vb\kpar\lt[(\pperp-\ppar)\frac{\dB}{B} + \dppar\rt]\\
&&-\, \kpar\lt(\pperp-\ppar + \frac{B^2}{4\pi}\rt)\dvb + \vk\cdot\delta\vG.
\label{eq:du}
\eea
Note that $\delta\vS = i\pperp(\vk\dV + \dV\vk)$,
where $\dV = \dvu + (\qperp\dvb + \vb\dqperp)/\pperp$.
Therefore, from \eqref{eq:G},
\bea
\nonumber
\lefteqn{\vk\cdot\delta\vG = \frac{i\pperp}{\Omega}\bigg\{
\lt(\kpar^2 + \frac{\kperp^2}{4}\rt)(\vb\times\dV)}&&\\ 
\nonumber
&&+\Bigl[\lt(\kpar\vb + \frac{\vkperp}{4}\rt)\lt(\vkperp\times\vb\rt)
- \lt(\vkperp\times\vb\rt)\lt(\kpar\vb + \frac{\vkperp}{4}\rt)
\Bigr]\cdot\dV\bigg\}\\
&&+\,\frac{1}{\Omega}\lt[\kpar\lt(\delta\vsig\times\vb\rt) 
+ \vkperp\cdot\lt(\delta\vsig\times\vb\rt)\vb\rt],\\
\lefteqn{\delta\vsig = -i\lt[(\pperp-\ppar)\lt(\omega\dvb - \kpar\dvu\rt)
-(3\qperp-\qpar)\kpar\dvb\rt].}&&
\eea 
In the above equations, $\dB=\dBpar$ and $\dvb=\dvBperp/B$, where 
$\dvB$ satisfies the perturbed \eqref{eq:ind}:
\bea
\nonumber
\lefteqn{\omega\,\frac{\dvB}{B} = -\kpar\dvuperp + \vb\lt(\vkperp\cdot\dvuperp\rt)}&&\\
\label{eq:dB}
&&+\,\,\frac{i\kpar}{mn\,\Omega}
\bigg\{\lt(\pperpe-\ppare + \frac{B^2}{4\pi}\rt)\lt(\vk\times\dvb\rt)\\
&&+\,\,\lt(\vkperp\times\vb\rt)
\lt[\dpperpe - \dppare -\lt(\pperpe - \ppare - \frac{B^2}{4\pi}\rt)
\frac{\dB}{B}\rt]
\bigg\}.
\nonumber
\eea

Examining \eqsdash{eq:du}{eq:dB}, we observe that in the simplest case 
of $\vkperp=0$, the Alfv\'enically polarized perturbations decouple from 
the compressive/slow-wave-polarized perturbations  
($\dn$, $\dB$, $\dupar$, $\dpperp$, $\dppar$, $\dqperp$ and $\dqpar$). 
No kinetic physics is required to study 
the stability of Alfv\'enic perturbations, which satisfy
\bea
\label{eq:duperp}
\lefteqn{mn \omega\dvuperp = -\kpar\lt(\pperpi-\ppari + \pperpe -\ppare
+ \frac{B^2}{4\pi}\rt)\dvb}&&\\
\nonumber
&&+\,\frac{i\kpar^2}{\Omega}\,
\vb\!\times\!\lt[\ppari\dvuperp + (\pperpi-\ppari)\,\frac{\omega}{\kpar}\,\dvb 
- (2\qperpi-\qpari)\dvb\rt]\!,\\
\lefteqn{\omega\dvb = -\kpar\dvuperp +
\frac{i\kpar^2}{mn\,\Omega}\lt(\pperpe-\ppare + \frac{B^2}{4\pi}\rt)
\lt(\vb\times\dvb\rt),}&&
\label{eq:db}
\eea
where we have restored species indices on pressures and heat fluxes; note 
that only ion FLR terms are kept in \eqref{eq:duperp}.
In the absence of FLR effects, \eqsand{eq:duperp}{eq:db} 
describe Alfv\'en waves with propagation 
speed modified by the pressure anisotropy. When $\pperp-\ppar<-B^2/4\pi$, 
it gives rise to the well known firehose instability with a growth rate 
$\gamma\propto\kpar$ \citep{Rosenbluth,Chandra,Parker,Vedenov}. 
The FLR gives rise to a dispersive correction that 
sets the wavenumber of the fastest-growing mode \citep{Kennel,Davidson}, 
but it also contains 
a contribution from the heat fluxes, which lead to a new instability. 

Let us combine \eqsand{eq:duperp}{eq:db} and 
non-dimensionalize everything:
\bea
\nonumber
\lefteqn{\obar^2 \dvb = \frac{k^2}{2}\!\lt(\Delta + \frac{2}{\beta}\rt)\dvb
+ \frac{ik^2}{2}\bigl[(1-\DD)\obar + k \gT\bigr]\lt(\vb\times\dvb\rt),}&&\\
&&
\label{eq:evproblem}
\eea
where $\obar = \omega/\Omega$, $k=\kpar\rho$, $\rho=\vth/\Omega$, 
$\vth=(2\ppari/mn)^{1/2}$.
The problem has four physical dimensionless parameters
\bea
\lefteqn{\Delta = \frac{\pperpi-\ppari + \pperpe-\ppare}{\ppari},\quad
\beta = \frac{8\pi\ppari}{B^2},}&&\\ 
\lefteqn{\DD = \frac{\pperpi-\ppari - (\pperpe-\ppare)}{\ppari} - \frac{2}{\beta},\quad 
\gT = \frac{2\qperpi-\qpari}{\ppari\vth},}&& 
\eea
but, in fact, only two matter because only the combination 
$\Delta+2/\beta$ figures in \eqref{eq:evproblem} 
and $\DD$ will turn out not to be of much consequence.
The resulting dispersion relation is
\beq 
\lt[\obar^2 - \frac{k^2}{2}\lt(\Delta + \frac{2}{\beta}\rt)\rt]^2 
= \frac{k^4}{4}\bigl[(1-\DD)\,\obar + k\gT\bigr]^2.
\eeq
This has four roots of which two can be unstable:
\bea
\nonumber
\obar &=& \pm\frac{k^2}{4}\,(1-\DD)\\ 
&&+\, \frac{i|k|}{\sqrt{2}}\sqrt{-\lt(\Delta + \frac{2}{\beta}\rt)
\mp k\gT - \frac{k^2}{8}(1-\DD)^2}
\label{eq:omega}
\eea
(we will henceforth refer to the positive/negative frequency 
modes as ``$+$/$-$ modes''). 
The instability occurs for $k$ such that 
the expression under the square root is positive. 
Demanding that the interval of such wavenumbers is non-empty 
gives the necessary and sufficient condition for instability:
\beq
\Lambda \equiv \gT^2 - \frac{(1-\DD)^2}{2}\lt(\Delta+\frac{2}{\beta}\rt)>0.
\label{eq:Lambda}
\eeq

\subsection{Firehose instability}

\begin{figure}
\includegraphics[width=84mm]{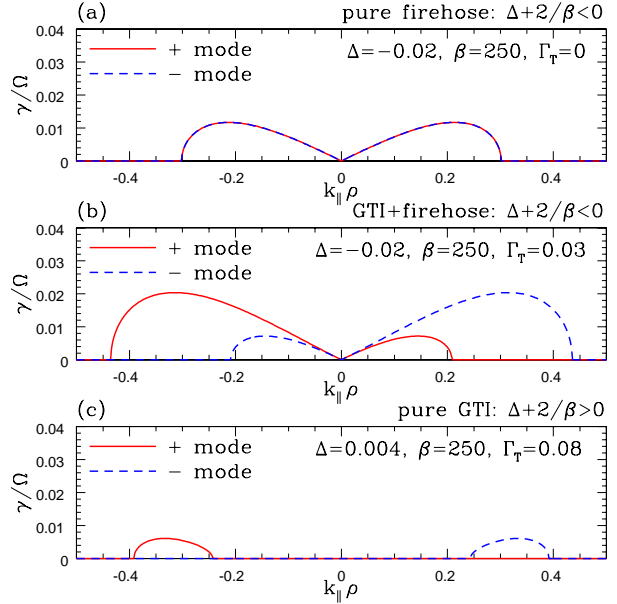}
\caption{Growth rates of the firehose and gyrothermal instabilities 
(\eqref{eq:omega})
in three qualitatively different regimes: 
(a) pure firehose, $\gT=0$; 
(b) GTI combined with firehose, $\Delta + 2/\beta<0$; 
(c) pure GTI, $\Delta + 2/\beta>0$ (firehose stable).
We have set $\DD=\Delta-2/\beta$, i.e., $\pperpe-\ppare=0$.}  
\label{fig:gamma}
\end{figure}

We observe first that if the heat fluxes are negligible, 
$\gT^2\ll |\Delta+2/\beta|$, this condition is satisfied for 
$\Delta+2/\beta<0$ and we have the standard parallel ($\kperp=0$) 
firehose dispersion relation \citep{Kennel,Davidson}:
\bea
\lefteqn{\obar = \pm\frac{k^2}{4}\,(1-\DD) + 
\frac{i|k|}{\sqrt{2}}\lt|\Delta+\frac{2}{\beta}\rt|^{1/2}\sqrt{1 - \frac{k^2}{k_0^2}},}&&\\
\lefteqn{k_0 = \frac{2\sqrt{2}}{|1-\DD|}\lt|\Delta + \frac{2}{\beta}\rt|^{1/2},}&&
\label{eq:k0}
\eea
where $k_0$ is the cutoff wavenumber and each of the $+$ and $-$ modes 
has two peaks of the growth rate occurring symmetrically at $\kp=\pm k_0/\sqrt{2}$
(see \figref{fig:gamma}{a}), where
\beq
\label{eq:gmax_firehose}
\gmax = \frac{1}{|1-\DD|}\lt|\Delta + \frac{2}{\beta}\rt|.
\eeq 
Note that here and everywhere else, we assume that $\Delta$ is not too close to $1$. 

\subsection{Gyrothermal instability}

The situation becomes more complicated when the heat fluxes are not negligible. 
Let us assume, without loss of generality, that $\gT>0$ (otherwise, change the sign 
of the parallel spatial coordinate). There are two unstable intervals: 
\bea
\label{eq:plus}
+~{\rm mode:}&&
-\frac{4\lt(\gT+\sqrt{\Lambda}\rt)}{(1-\DD)^2}<k<-\frac{4\lt(\gT-\sqrt{\Lambda}\rt)}{(1-\DD)^2},\\
-~{\rm mode:}&&
~~\frac{4\lt(\gT-\sqrt{\Lambda}\rt)}{(1-\DD)^2}<k<\frac{4\lt(\gT+\sqrt{\Lambda}\rt)}{(1-\DD)^2}.
\label{eq:minus}
\eea
When $\Delta+2/\beta\le0$, these intervals intersect and contain $k=0$, 
otherwise they are disjoint (see \figref{fig:gamma}{b,c}). 
Computing their peak growth rates and corresponding 
wavenumbers is straightforward. Here we consider two interesting limits. 

When $\gT^2\gg|\Delta+2/\beta|$, we have, for the $+$ and $-$ modes respectively: 
\beq
\label{eq:gmax_gti}
\kp = \mp\frac{6\gT}{(1-\DD)^2},\quad
\gmax = \frac{3\sqrt{3}\,\gT^2}{|1-\DD|^3}.
\eeq
We see that an instability is present that is driven purely by 
heat fluxes, even when the pressure anisotropy is neutralized 
by the tension force ($\Delta=-2/\beta$). This is the purest form of 
the {\em gyrothermal instability (GTI)}, which, as far as we know, 
has not been previously reported in the literature. In the more general 
case when the pressure anisotropy is not negligible, the GTI 
operates in conjunction with the firehose. 
The condition \exref{eq:Lambda} means that GTI can be operative even 
when $\Delta+2/\beta>0$, a regime in which the Alfv\'en waves 
have previously been believed to be stable. 

The second important limit is the case when GTI is close to marginal stability, 
$\Lambda\to+0$ (we are assuming that $\gT^2$ is finite, so the firehose is stable 
in this limit). According to \eqsand{eq:plus}{eq:minus}, the instability intervals 
in this limit shrink to the immediate vicinity of just two wavenumbers:
\beq
\label{eq:GTImarg}
\kp = \mp\frac{4\gT}{(1-\DD)^2}\lt(1+\frac{\Lambda}{\gT^2}\rt),\quad
\gmax = \frac{4\gT\sqrt{\Lambda}}{|1-\DD|^3},
\eeq
where the upper sign is for the $+$ mode, the lower for the $-$ mode.
This is a very different behaviour from the firehose, for which 
the interval of growing modes moves to ever longer wavelengths 
as marginal stability ($\Delta+2/\beta\to-0$) is approached 
(see \eqref{eq:k0}), i.e., the firehose stops being a microscale 
instability in this limit. In contrast, the GTI always excites 
Alfv\'enic fluctuations at very short wavelengths. 

Finally, we 
note that the assumption in our derivation that 
$\omega/\Omega\ll1$ and $\kpar\rho\ll1$ imposes constraints
on the values of our dimensionless parameters that we are 
allowed to consider: $|\Delta+2/\beta|\ll1$ for the firehose 
and $\gT\ll1$ for the GTI. The expressions for maximum growth rates and 
corresponding wavenumbers derived above 
(\eqsref{eq:gmax_firehose}, \exref{eq:gmax_gti} and \exref{eq:GTImarg})
provide guidance on the relative smallness of all these quantities 
and, therefore, on the ordering schemes that can be pursued in 
weakly nonlinear theories (one example is the ordering 
adopted by \citealt{Rosin}). 

\section{Nonlinear evolution}
\label{sec:nlin}

Nonlinear theories of pressure-anisotropy-driven plasma instabilities 
are in their infancy, but most of them agree that the net result 
is to drive the anisotropies towards marginal stability thresholds 
\citep[e.g.,][]{Shapiro,Quest,Matteini06,Schek08,Califano,Istomin,Rosin}. 
Observational evidence from the solar wind strongly points in the 
same direction \citep{Kasper,Hellinger,Matteini07,Bale}.

If we assume that this is what happens 
in the case of the firehose and gyrothermal instabilities, 
then the marginal state $\Lambda=0$ 
(see \eqref{eq:Lambda}) implies a certain relationship between 
the heat fluxes and the pressure anisotropy in the nonlinear regime. 
In order to find the way in which the system contrives to 
set up this relationship, we must first examine the physical 
mechanisms that determine $\Delta$, $\qperp$ and~$\qpar$. 

Subtracting \eqref{eq:ppar} from \eqref{eq:pperp}, we get 
\bea
\nonumber
\lefteqn{\frac{\did(\pperp-\ppar)}{\did t} = (\pperp+2\ppar)\frac{1}{B}\frac{\did B}{\did t}
-(3\ppar-\pperp)\frac{1}{n}\frac{\did n}{\did t}}&&\\ 
&&\quad\qquad-\, \vdel\cdot\lt(\vqperp-\vqpar\rt) - 3\qperp\vdel\cdot\vb - 3\nu(\pperp-\ppar).
\label{eq:aniso}
\eea
This tells us that there are three sources of pressure anisotropy: 
changing magnetic-field strength (changes 
in $\pperp$ have to match changes in $B$ to maintain conservation 
of the first adiabatic invariant for each particle, $\mu=m\vperp^2/2B$), 
compression/rarefaction, and heat fluxes.

If we assume for a moment that the collision rate 
is larger than the rate of change of all fields, then
the differences between $\pperp$ and $\ppar$ in \eqref{eq:aniso} can 
be neglected everywhere except the collisional term and so 
the steady-state average pressure anisotropy satisfies
\bea
\nonumber
\lefteqn{\Delta = 
\frac{1}{\nu}\Biggl\{
\frac{1}{B}\frac{\did B}{\did t}
-\frac{2}{3}
\frac{1}{n}\frac{\did n}{\did t}
-\frac{\vdel\cdot\lt[\vb(\qperp-\qpar)\rt] + 3\qperp\vdel\cdot\vb}{3\ppar}\Biggr\}.}&&\\
&&\label{eq:aniso_coll}
\eea
Note that if we use \eqsand{eq:ind}{eq:cont} (neglecting FLR terms in the 
induction equation) to express the rates of change of $B$ and $n$ 
in the right-hand side of \eqref{eq:aniso_coll},
the first two terms are the \citet{Braginskii} 
parallel viscous stress. The last term is the heat-flux correction 
to it introduced by \citet{Mikh1,Mikh2} for subsonic flows.
Under the same assumption of high collisionality, the heat fluxes 
are\footnote{The numerical prefactor in the last expression in 
\eqref{eq:q_coll} depends on the exact form of the collision operator used 
and is not relevant to our discussion. The same applies to numerical coefficients 
in \eqref{eq:aniso_coll} and so, to preserve consistency, we have given the values 
obtained by using the Lorentz operator \citep{Rosin}. 
The more precise coefficients for ions are $25/32$ in \eqref{eq:q_coll}, 
$3075/1068$ in front of the first two terms in \eqref{eq:aniso_coll}, 
and $1823/1068$ in front of the heat flux terms in the same equation; 
the ion collision frequency is $\nu=4\sqrt{\pi}ne^4\Lambda/3m^{1/2}T^{3/2}$, 
where $\Lambda$ is the Coulomb logarithm \citep{Braginskii,Mikh1,Mikh2,Catto}.} 
\beq
\qperp = \frac{1}{3}\,\qpar = 
-\frac{1}{2}\,n\,\frac{\vth^2}{\nu}\,\vb\cdot\vdel T,
\label{eq:q_coll}
\eeq
where $T=p/n$ and $p=(2/3)\pperp+(1/3)\ppar$. 

As we showed in \secref{sec:lin}, the slow macroscale motions that 
produce this $\Delta$ and these heat fluxes are unstable to microscale 
perturbations, in particular, the Alfv\'enic ones  
excited by the firehose/GTI. \citet{Schek08} showed 
that the way a sea of small-scale Alfv\'enic fluctuations 
can change a large-scale driven anisotropy is by growing secularly 
with time and thus producing a finite change in 
the {\em average} field strength: 
\beq
\overline{\frac{1}{B}\frac{\did B}{\did t}} = 
\frac{1}{B_0}\frac{\did B_0}{\did t} 
+ \frac{1}{2}\frac{\dd\overline{|\dvb|^2}}{\dd t},
\label{eq:avgdB}
\eeq 
where the overbar denotes a small-scale average, 
$B_0$ is the slowly changing macroscale field 
and $\dvb=\dvBperp/B_0$ is the fast microscale 
Alfv\'enic perturbation of it. Let us replace the magnetic term 
in \eqref{eq:aniso_coll} with its average given by \eqref{eq:avgdB}.
Even though the fluctuation amplitude is small, 
the nonlinear feedback will produce a finite contribution to $\Delta$ 
if the fluctuation energy grows secularly, 
$\overline{|\dvb|^2}\sim\gamma_0 t$, where 
$\gamma_0$ is the typical rate of change of $B_0$. 
There does not appear to be any other way 
for the small Alfv\'enic fluctuations to affect the average 
macroscopic pressure anisotropy or heat fluxes. 

In the case of the pure firehose instability (no heat fluxes), 
the nonlinear feedback described above cancels the negative pressure 
anisotropy that triggered the firehose and pushes the system towards 
$\Delta + 2/\beta\to-0$. If heat fluxes are present, the marginal 
state of the GTI requires $\Delta + 2/\beta>0$ ($\Lambda\to+0$; 
see \eqref{eq:Lambda}). This can still be achieved by 
secularly growing Alfv\'enic fluctuations (which, unlike for the 
firehose, now have a definite scale unaffected by the pressure 
anisotropy; this is explored further in \citealt{Rosin}). 

A remarkable consequence of this predicted tendency for 
a system to develop positive pressure anisotropy to cancel 
the destabilizing effect of heat fluxes is that instabilities 
associated with $\Delta>0$ (e.g., mirror) could perhaps be triggered  
as secondary instabilities of the saturated state of 
the GTI. One might imagine a sea of Alfv\'enic fluctuations attempting 
to neutralize the GTI and exciting unstable mirror modes --- this 
is feasible if the pressure anisotropy corresponding to the marginal 
state of the GTI exceeds the mirror stability threshold: 
$\Delta\simeq2\gT^2-2/\beta>1/\beta$, i.e., $\gT^2>3/2\beta$.
The mirror mode near its threshold $\Delta-1/\beta\to+0$ 
is polarized as a highly oblique slow wave: it  
has $\dupar$ and $\dBpar$ with $\kpar\rho\sim\Delta-1/\beta\ll
\kperp\rho\sim(\Delta-1/\beta)^{1/2}$ \citep[see, e.g.,][]{Hellinger_mirror}. 
This suggests a three-scale system: 
a macroscale equilibrium, the microscale Alfv\'enic foam with 
$\kpar\rho\sim\gT\sim1/\sqrt{\beta}$ (see \eqref{eq:GTImarg}) 
driven by the GTI and producing an average pressure anisotropy, 
and a mesoscale near-threshold mirror turbulence driven by that anisotropy
and, because of scale separation, probably otherwise disconnected 
from the Alfv\'enic modes. 
Finding out how they all coexist and how the mirror saturates requires 
a systematic kinetic calculation, which will be attempted elsewhere.  

Finally, as an alternative to the above considerations, 
we should perhaps mention the possibility 
of strong nonlinear distortions of the magnetic field ($\dvb\sim1$) 
that could reorient the field so as to minimize 
the parallel ion temperature gradient and thus switch off 
or weaken the GTI --- on large scales, 
such behaviour has been observed in simulations of another, 
macroscale, instability driven by the parallel (electron) heat flux 
and buoyancy force, called the heat-flux-buoyancy instability, or HBI 
\citep{Sharma_HBI,Parrish_HBI,Bogdanovic}.

\section{Physical and astrophysical considerations}
\label{sec:disc}

\subsection{Physical conclusions}
\label{sec:physconc}

The main physical conclusion is that parallel heat fluxes 
can directly drive microscale instabilities in magnetized 
astrophysical plasmas. This can happen in two ways. 

First, as follows from \eqref{eq:aniso}, plasma 
pressure anisotropy can be driven by heat fluxes, so firehose, mirror 
and the rest of the microinstabilities due to $\pperp-\ppar\neq0$ 
can be triggered not just by plasma motions, but also by parallel 
temperature gradients. Although perhaps not much discussed explicitly, 
this instability mechanism is not particularly surprising and 
it is implicitly present in the existing analytical and numerical 
models based on CGL equations with heat fluxes 
\citep[e.g.,][]{Snyder1,Quataert_kinMRI,Sharma06,Sharma07}. 

A more interesting and, we believe, novel instability mechanism 
is the destabilization of the Alfv\'enic perturbations by the 
ion parallel heat fluxes via the FLR effects in the plasma 
pressure tensor --- we call this the gyrothermal instability (GTI). 
When the firehose is unstable, the GTI can substantially 
modify (increase) its growth rate, but more importantly, 
the GTI persists even when the firehose is stable, so the 
firehose marginal stability condition has to be replaced 
by the GTI marginal stability condition involving both the 
pressure anisotropy and the ion heat flux (\eqref{eq:Lambda}). 

The GTI is distinct 
from the two other instabilities associated with the presence 
of temperature gradients and recently explored in astrophysical contexts 
--- the MTI \citep{Balbus_MTI,Parrish_MTI1,Parrish_MTI2} and 
the HBI \citep{Quataert_HBI,Sharma_HBI,Parrish_HBI,Bogdanovic,Ruszkowski}. 
The latter are driven by buoyancy and are essentially 
macroscale fluid instabilities, like MRI \citep{Balbus_review} 
or MVI \citep{Balbus_MVI}. They are also much slower than the 
GTI, which is a microscale plasma instability belonging to the 
same class as the firehose, 
with peak growth rate a fraction of the cyclotron frequency. 
Since such an instability can be triggered 
by the presence of a heat flux, one might wonder whether in the same 
way that large-scale pressure anisotropy could be conjectured 
always to be determined by the marginal stability conditions 
of the microinstabilities \citep{Sharma06,Sharma07,Schek06,Lyutikov,Kunz}, 
the heat fluxes as well should be constrained by the marginal stability 
conditions of the GTI and, perhaps, other such instabilities. 
We stress, however, that, whereas this might be a reasonable interim 
course of action, it by no means excuses us from the task of finding 
out how GTI and the rest of the instabilities behave and saturate 
on the microphysical level (see discussion in \secref{sec:nlin}).  

\subsection{An astrophysical example: galaxy clusters}
\label{sec:cores}

A detailed development of applications to concrete astrophysical 
systems falls outside the scope of this paper \citep[see, e.g.,][]{Kunz}. 
However, it is, 
perhaps, illuminating to provide a few estimates of the role 
the GTI might play in cool cores of relaxed 
galaxy clusters, a good example of a real astrophysical plasma 
for which a sufficient amount of observational evidence exists 
to enable a quantitative discussion of the multiscale dynamics. 

\subsubsection{Three-scale dynamics}
\label{sec:threescale}

The conditions in the cluster cores 
are believed to be controlled by a balance between the radiative 
cooling and a reheating due perhaps to electron heat conduction from 
the bulk of the cluster and perhaps also to the turbulence 
excited by the active galactic nuclei 
\citep[e.g.,][and references therein]{Binney,Dennis,Peterson,McNamara,Guo,Ruszkowski}. 
The plasma in the cores has the electron density $n_e$ in the range 
$10^{-2}$ to $10^{-1}$ cm$^{-3}$ at the radial distance of 
$r\sim10$ kpc from the centre and about a 
factor of 10 less at the edge of the core at $r\sim 100$ kpc.
The ion density is the same for a hydrogen plasma. 
The electron temperature $T_e$ is measured reasonably precisely 
and is of the order of a few keV, rising by about a factor of 
2 or 3 from $r\sim10$ kpc to $100$ kpc 
\citep[e.g.,][]{David,Vikhlinin,Fabian,Leccardi,Sanders1,Sanders2}. 
The ion temperature is not 
measured, but the ion-electron temperature equilibration 
turns out to be quite fast compared to all other relevant 
dynamics, so $T_i\sim T_e$ can reasonably be assumed. 
The unsolved macroscale problem is why the temperature does not 
drop lower in the centre --- simple estimates suggest that 
the system should be vulnerable to a collapse onto the centre 
precipitated by the radiative cooling on a characteristic 
time scale of about 1~Gyr. 

This is where turbulent heat conduction\footnote{Since the cooling rate 
is $\propto n_e T_e^{-1/2}$ and the relaxation rate of temperature 
gradients based on Spitzer conductivity is $\propto n_e^{-1} T_e^{5/2}$ \citep{Spitzer}, 
they cannot balance in a stable way, so Spitzer conduction by itself 
is not sufficient to explain the absence of the cooling catastrophe.
In contrast, turbulent heating controlled by the plasma instabilities 
via pressure anisotropy (as explained below) turns out to be a thermally 
stable mechanism for regulating cooling flows \citep{Kunz}.} 
and turbulent heating are invoked as mechanisms that prevent 
the cooling catastrophe. The outer scale $L$ of turbulent 
motions is believed to be between a few and a few tens of kpc, 
with corresponding velocities 
$U$ of a few hundred ${\rm km\,s}^{-1}$ \citep{Ensslin,Sanders2}.
The turbulent motions lead to fluctuations in the magnetic-field 
strength and so excite pressure anisotropies, given by \eqref{eq:aniso_coll}. 
\Eqref{eq:ind} tells us that the typical rate of change of the field 
is comparable to the typical rate of strain $\sim (U/L)\Re^{1/2}$, where 
$\Re\sim UL\nu/\vth^2$ is the Reynolds number (the maximum 
rate of strain that can affect the magnetic-field strength 
is at the viscous scale set by the parallel viscosity; see 
\citealt{Schek06} for a detailed explanation). Thus, 
we estimate the pressure anisotropy as follows:
\bea
\nonumber
\Delta&\sim&
\frac{1}{\nu}\frac{U}{L}\,\Re^{1/2}
\sim \lt(\frac{1}{\nu}\frac{U}{L}\rt)^{1/2}\frac{U}{\vth}\\
\nonumber
&\sim&
0.007\lt(\frac{n_e}{0.01\,\rm{cm}^{-3}}\rt)^{-1/2}
\lt(\frac{T_i}{1\,\rm{keV}}\rt)^{1/4}\\
&&\times
\lt(\frac{U}{100\,\rm{km\, s}^{-1}}\rt)^{3/2}
\lt(\frac{L}{10\,\rm{kpc}}\rt)^{-1/2},
\eea
where $\nu$ is the ion collision rate. 
In view of the instability condition \exref{eq:Lambda}, 
whether this anisotropy will trigger plasma microinstabilities is decided 
by comparing it with 
\bea
\frac{2}{\beta} = 0.005 \lt(\frac{B}{1\,\mu\rm{G}}\rt)^2
\lt(\frac{n_e}{0.01\,\rm{cm}^{-3}}\rt)^{-1}
\lt(\frac{T_i}{1\,\rm{keV}}\rt)^{-1}.
\eea
The two numbers are remarkably close (obviously, only orders of magnitude 
matter here, given all the uncertainties). Thus the intracluster 
plasma teeters at the brink of marginal stability. 
In the unstable state, at the reference values $B=1\,\mu$G and $T_i=1\,$keV, 
the firehose (or GTI) will have growth times and peak-growth scales 
\bea
\lefteqn{\gmax^{-1} \sim \lt(\Delta\Omega_i\rt)^{-1} 
\sim 2\cdot10^4\,\rm{s}\simeq 6\,\rm{hr},}&&\\
\lefteqn{\kp^{-1} \sim \Delta^{-1/2}\rho_i \sim 700,000\,\rm{km}
\simeq 20\,\rm{npc}}&&
\eea
(see \secref{sec:lin}). 
These are microscopic scales compared both to global 
cluster dynamics and intracluster turbulence. The implication is that 
the plasma instabilities should saturate and presumably contrive to return the 
intracluster medium to marginal stability instantaneously fast via 
an observationally invisible sea of nanoparsec-scale magnetic fluctuations. 

Thus, a cluster core is a ``three-scale system'': global
equilibrium profiles ($10^2$\,kpc, $10^0$\,Gyr) and  
turbulence ($10^1$\,kpc, $10^1$\,Myr) constitute the macroscale 
magnetofluid dynamics of the intracluster medium,\footnote{As we already 
pointed out in \secref{sec:physconc}, 
various macroscopic instabilities that play an 
important part in plasma dynamics, including those due to 
plasma effects such as anisotropic viscosity 
and thermal conductivity (MVI, MTI, HBI) 
act on time scales roughly comparable with the turbulence 
and are slow compared to the microinstabilities: e.g., 
HBI in cluster cores is estimated to have growth times 
of order $10^2$\,Myr \citep{Parrish_HBI}.} 
subject to transport properties controlled by ``nanoscales'' 
($10^1$\,npc, $10^1$\,hr), where plasma microinstabilities are excited. 
Their nonlinear behaviour sets the pressure anisotropy and 
probably also the heat fluxes. 
The pressure anisotropy determines the effective viscosity 
of the plasma and, therefore the heating rate; the heat 
fluxes determine the effective thermal conductivity --- thus, 
neither the turbulence nor the global dynamics (e.g., temperature 
profiles for the cooling-core problem) can 
be computed correctly without a good theory or, at least, a good 
model prescription, for the effect of the microinstabilities 
on the macroscale dynamics.  
A similar three-scale situation arises in most other weakly 
collisional\footnote{Collisional scales are intermediate between turbulence 
and plasma microphysics: the collision times are 
$\nu_{ii}^{-1}\sim 0.04\,$Myr, $\nu_{ei}^{-1}\sim0.001\,$Myr, 
$\nu_{ie}^{-1}\sim1$\,Myr (the latter is the typical time for $T_i$ and $T_e$ 
to equalize); the mean free path is $\mfp\sim0.01\,$kpc, where we have 
taken reference values of $n_e=0.01\,\rm{cm}^{-3}$ and $T_i=1\,$keV, 
collision frequencies are $\propto n T^{-3/2}$ and 
$\mfp\propto n^{-1}T^2$.} 
cosmic plasmas: e.g., accretion flows, solar wind, etc. 

\subsubsection{Temperature fluctuations}
\label{sec:tempfluct}

As we have shown in this paper, ion temperature gradients, including 
ones due to temperature fluctuations, if they are there and if the 
associated parallel heat fluxes are large enough, will excite 
microinstabilities. The estimates of $\gmax$ and $\kp$ in \secref{sec:threescale} 
still hold, by order of magnitude, for the GTI, so the instability is 
extremely fast and one should expect to find plasma close to the marginal 
state. We may estimate (crudely), 
the minimum parallel temperature length scale allowed by the 
instability condition \exref{eq:Lambda} by requiring 
$\gT^2 \lesssim 2/\beta$ for stability and using \eqref{eq:q_coll} 
for the heat fluxes:
\beq
l_T \gtrsim 0.3 \lt(\frac{n_e}{0.01\,\rm{cm}^{-3}}\rt)^{-1/2}
\lt(\frac{T_i}{1\,\rm{keV}}\rt)^{5/2}
\lt(\frac{B}{1\,\mu\rm{G}}\rt)^{-1}\rm{kpc},
\label{eq:lT}
\eeq
where $l_T^{-1} = \vb\cdot\vdel\ln T$ is the temperature scale. 
Note the very strong temperature dependence 
of this lower bound: thus, deep in the cool cores, the estimate 
above gives kiloparsec-scale temperature fluctuations, 
rising to tens and even hundreds of kpc at larger distances 
from the centre. 

Interestingly, temperature fluctuations on 1 to 10 kpc scales 
have been detected in cool-core clusters \citep{Simionescu,Fabian,Sanders1}
while in the bulk of the cluster gas and in non-cool-core (radio-halo) 
clusters, the scales appear to be larger, around 100\,kpc 
\citep{Markevitch,Million}. Thus, we again find the observed 
physical conditions intriguingly close to the marginal stability 
conditions set by plasma microphysics. Nevertheless, we would like to 
conclude on a cautious note: whether the plasma contrives to satisfy the 
lower bound \exref{eq:lT} by smoothing the temperature gradients or by 
aligning them carefully across the magnetic field remains unclear and 
underscores the need for a detailed theory of the nonlinear saturation 
of the GTI and other plasma microinstabilities. Observationally, it 
would be fascinating to see if any evidence can be obtained 
of correlations between the magnetic field direction and temperature 
fluctuations --- presumably not an impossible task if one combines 
radio observations of polarized synchrotron emission and X-ray 
temperature maps \citep[cf.][]{Taylor}. 

\section*{Acknowledgments}
It is a pleasure to thank S.~Balbus, J.~Binney, G.~Hammett, 
M.~Kunz and J.~Stone for discussions and M.~Markevitch for pointing out 
some of the observational evidence discussed in \secref{sec:tempfluct}. 
This work was supported by STFC (AAS \& MSR), 
and the Leverhulme Trust Network for Magnetized Plasma Turbulence (FR).

\appendix

\section{Plasma pressure tensor}
\label{app:P}

We start with the general kinetic equation for the distribution 
function of a plasma species: 
\beq
\frac{\did f}{\did t} + \vv\cdot\vdel f + 
\lt(\va + \frac{e}{m}\frac{\vv\times\vB}{c} - \vv\cdot\vdel\vu\rt)
\!\cdot\!\frac{\dd f}{\dd\vv}
= \St{f},
\label{eq:Vlasov}
\eeq
where $e$ is the particle charge, $c$ the speed of light, 
$\va = (e/m)(\vE + \vu\times\vB/c) - \did\vu/\did t$
and $C[f]$ is the collision integral.
The flow velocity $\vu$ appears in the kinetic equation because $\vv$ is the peculiar velocity. 
Since $(e/mc)(\vv\times\vB)\cdot\dd f/\dd\vv = -\Omega\dd f/\dd\vartheta$, 
where $\Omega$ is the Larmor frequency and $\vartheta$ is the phase angle 
of the particle's gyration around the magnetic-field line, 
\eqref{eq:Vlasov} can be rewritten as follows
\beq
\Omega\,\frac{\dd f}{\dd\vartheta} = 
\frac{\did f}{\did t} + \vv\cdot\vdel f + 
\lt(\va - \vv\cdot\vdel\vu\rt)\cdot\frac{\dd f}{\dd\vv}
- \St{f}.
\label{eq:df}
\eeq
We can now express the plasma pressure tensor $\vP=m\int\did^3\vv\,\vv\vv f$
using the following identity
\bea
\vv\vv &=& \frac{\vperp^2}{2}\lt(\vI - \vb\vb\rt) + \vpar^2\vb\vb
+ \frac{\dd\vT}{\dd\vartheta},\\
\nonumber
\vT &=& \lt(\vpar\vb + \frac{\vvperp}{4}\rt)\lt(\vvperp\times\vb\rt) 
+ \lt(\vvperp\times\vb\rt)\lt(\vpar\vb + \frac{\vvperp}{4}\rt),
\eea
or, in index notation, $T_{ij} = (1/4)M_{ijkl}v_k v_l$, where 
\beq
M_{ijkl} = \lt(\delta_{ik} + 3b_i b_k\rt)\eps_{jln} b_n 
+ \eps_{iln}b_n\lt(\delta_{jk} + 3b_j b_k\rt).
\eeq
Therefore, after integration by parts with respect to $\vartheta$, 
\beq
P_{ij} = \pperp\delta_{ij} - (\pperp-\ppar) b_i b_j 
- \frac{M_{ijkl}}{4}\!\int\!\did^3\vv\,m v_k v_l\,\frac{\dd f}{\dd\vartheta}.
\eeq
We now substitute \eqref{eq:df} into the above expression 
and notice that $\int\did^3\vv\,\vv\vv\,\va\cdot\dd f/\dd\vv = 0$ 
after integration by parts and using the fact that 
$\int\did^3\vv\,\vv f=0$ by definition of peculiar velocity. 
We get, therefore,
\bea
\nonumber
P_{ij} &=& \pperp\delta_{ij} - (\pperp-\ppar) b_i b_j 
- \frac{M_{ijkl}}{4\Omega}\biggl[\frac{\did P_{kl}}{\did t} + \nabla_m Q_{mkl}\biggr.\\
&&+\biggl.\bigl(\delta_{mn} P_{kl} + \delta_{kn}P_{ml} + \delta_{ln}P_{mk}\bigr)\nabla_m u_n
- C_{kl}\biggr],
\label{eq:Pij}
\eea
where $C_{kl}= m\int\did^3\vv\,v_k v_l\St{f}$ and we have introduced 
the heat flux tensor $Q_{mkl} = m\int\did^3\vv\,v_m v_k v_l f$. 

So far we have made no approximations.
As promised in \secref{sec:eqns}, 
we now calculate all terms in \eqref{eq:Pij} 
assuming that we can use a gyrotropic (independent of $\vartheta$) 
distribution function. This amounts to setting up a perturbation 
theory in which to lowest order, \eqref{eq:df} gives a gyrotropic 
equilibrium distribution, $\Omega\dd f_0/\dd\vartheta = 0$, and at the 
next order we have $\Omega\dd\delta f/\dd\vartheta = \dots$, where 
only $f_0$ appears in the right-hand side. The assumptions we need 
to achieve such an expansion are $\omega/\Omega\ll1$ and $k\rho\ll1$
for all quantities involved. 

Since $f_0$ is gyrotropic, 
we may gyroaverage $\lt<v_k v_l\rt> = 
(1/2\pi)\int\did\vartheta\,v_k v_l = 
(\vperp^2/2)(\delta_{kl}-b_k b_l) + \vpar^2 b_k b_l$
inside all the velocity integrals 
in the square brackets in \eqref{eq:Pij}, so we get
\bea
\label{eq:MPkl}
\lefteqn{M_{ijkl}P_{kl}\nabla_m u_m = M_{ijkl}C_{kl} = 0,}&&\\
\nonumber
\lefteqn{M_{ijkl}P_{ml}\nabla_m u_k =}&& \\ 
&&-\pperp\lt[
\vb\times(\vdel\vu)\cdot\lt(\vI + 3\vb\vb\rt)
-\lt(\vI + 3\vb\vb\rt)\cdot(\vdel\vu)^T\times\vb 
\rt],\\
\nonumber
\lefteqn{M_{ijkl}P_{mk}\nabla_m u_l =}&& \\
\nonumber
&&-\pperp\lt[
\vb\times(\vdel\vu)^T\cdot\lt(\vI + 3\vb\vb\rt)
- \lt(\vI + 3\vb\vb\rt)\cdot(\vdel\vu)\times\vb 
\rt]\\
&&-4(\pperp-\ppar)\bigl[\vb\lt(\vb\cdot\vdel\vu\times\vb\rt) 
+ \lt(\vb\cdot\vdel\vu\times\vb\rt)\vb\bigr],\\
\lefteqn{M_{ijkl}\frac{\did P_{kl}}{\did t} = 
-4(\pperp-\ppar)\lt(\vb\,\frac{\did\vb}{\did t}\times\vb 
+ \frac{\did\vb}{\did t}\times\vb\,\vb\rt).}&&
\label{eq:Mdt}
\eea
Similarly gyroaveraging $\lt<v_mv_kv_l\rt>$ in the heat flux integral,
$Q_{mkl} =
\qperp\lt(b_m\delta_{kl} + \delta_{mk}b_l + \delta_{ml}b_k\rt)
- (3\qperp-\qpar)b_m b_k b_l$.
Therefore,
\bea
\nonumber
\lefteqn{M_{ijkl}\nabla_m Q_{mkl}=
\lt(\vI + 3\vb\vb\rt)\cdot\bigl[\vdel\vqperp + (\vdel\vqperp)^T\bigr]\times\vb}&& \\
\nonumber
&&-\,\vb\times\bigl[\vdel\vqperp + (\vdel\vqperp)^T\bigr]\cdot\lt(\vI + 3\vb\vb\rt)\\
&&-\,4(3\qperp-\qpar)\bigl[\vb\lt(\vb\cdot\vdel\vb\times\vb\rt)
+ \lt(\vb\cdot\vdel\vb\times\vb\rt)\vb\bigr].
\label{eq:MQ}
\eea
Assembling \eqsdash{eq:MPkl}{eq:Mdt} and \exref{eq:MQ} together 
in \eqref{eq:Pij}, we obtain \eqsdash{eq:Pfull}{eq:sig_def}. 

\section{CGL equations}
\label{app:KMHD}

In order to derive \eqsand{eq:pperp}{eq:ppar}, we average
\eqref{eq:df} over the gyroangles, $(1/2\pi)\int\did\vartheta$, 
which eliminates the left-hand side. In the remainder, we assume that 
the lowest-order distribution function is gyrotropic and so 
can be written as $f = f(t,\vr,v,\vpar)$. The time 
and spatial derivatives in \eqref{eq:df} are taken at constant 
$\vv$, so, in order for the gyroaverage to commute with them, 
they have to be transformed into derivatives at constant 
$v$ and $\vpar$, a nontrivial step because 
$\vpar=\vv\cdot\vb(t,\vr)$: 
\bea
\lt(\frac{\did f}{\did t}\rt)_{\vv} &=& 
\lt(\frac{\did f}{\did t}\rt)_{v,\vpar}
+ \frac{\did\vb}{\did t}\cdot\vv\lt(\frac{\dd f}{\dd\vpar}\rt)_{v},\\
\lt(\vdel f\rt)_{\vv} &=& \lt(\vdel f\rt)_{v,\vpar} 
+ \lt(\vdel\vb\rt)\cdot\vv\lt(\frac{\dd f}{\dd\vpar}\rt)_{v}.
\eea
Using these formulae and also 
$\dd f/\dd\vv = (\vv/v)\dd f/\dd v + \vb\dd f/\dd\vpar$
and $\lt<\vv\rt> = \vpar\vb$, 
$\lt<\vv\vv\rt> = (\vperp^2/2)(\vI-\vb\vb) + \vpar^2\vb\vb$, 
we find that the gyroaveraged \eqref{eq:df} is
\bea
\nonumber
\lefteqn{\frac{\did f}{\did t} + \vpar\vb\cdot\vdel f 
+ \frac{\vperp^2}{2}\lt(\vdel\cdot\vb\rt)\frac{\dd f}{\dd\vpar}
+ \va\cdot\vb\lt(\frac{\vpar}{v}\frac{\dd f}{\dd v} + \frac{\dd f}{\dd\vpar}\rt)}&&\\
\nonumber
&&+ \lt(\vb\vb:\vdel\vu\rt)\lt[\lt(\frac{\vperp^2}{2}-\vpar^2\rt)\frac{1}{v}\frac{\dd f}{\dd v} 
- \vpar\frac{\dd f}{\dd\vpar}\rt]\\ 
&&- \lt(\vdel\cdot\vu\rt)\frac{\vperp^2}{2}\frac{1}{v}\frac{\dd f}{\dd v}
= C[f].
\label{eq:KinMHD}
\eea
Changing variables from $(v,\vpar)$ to $(\vperp,\vpar)$ or $(\mu,\vpar)$, 
where $\mu=\vperp^2/2B$, transforms this equation into forms that are perhaps more 
familiar from the well-known Kinetic MHD 
approximation \citep{Kulsrud}. 

\Eqsand{eq:pperp}{eq:ppar} are obtained
by taking the $\vperp^2/2$ and $\vpar^2$ moments of \eqref{eq:KinMHD}
and integrating by parts wherever opportune. The collisional relaxation 
terms are easiest to calculate with a simplified collisional 
operator, e.g., Krook \citep{Snyder1} or Lorentz \citep{Rosin}. 
To complete the picture, it may be useful to mention here that 
in some cases, especially when the pressure anisotropy $\pperp-\ppar$
is small compared to the pressures themselves, it is convenient 
to replace \eqsand{eq:pperp}{eq:ppar} by \eqref{eq:aniso} determining 
the evolution of $\pperp-\ppar$ and an equation for the total 
pressure $p=(2/3)\pperp+(1/3)\ppar$ or temperature $T$ defined by $p=nT$. 
Using \eqsand{eq:pperp}{eq:ppar}, we get
\beq
\frac{3}{2}\,n\,\frac{\did T}{\did t} 
= p\,\frac{1}{n}\frac{\did n}{\did t} 
+ (\pperp-\ppar)\lt(\frac{1}{B}\frac{\did B}{\did t}
-\frac{2}{3}\frac{1}{n}\frac{\did n}{\did t}\rt) 
- \vdel\cdot\vq,
\eeq
where $\vq = \vqperp+\vqpar/2$. 
The first term is compressional heating, the second 
viscous heating and the third the heat flux. 
While the same-species collisions do not affect 
the evolution of temperature (because of the energy and particle 
conservation), we do have to add to the above equation 
a temperature equilibration term, $-(3/2)n_i\nu_{ie}(T_i-T_e)$ for ions 
and negative of the same for electrons, where $\nu_{ie}$ is the ion-electron 
collision frequency (the ion-electron temperature 
equilibration terms were omitted in \eqsand{eq:pperp}{eq:ppar} because 
the relaxation of the pressure anisotropy was the dominant collisional 
effect there). In situations where radiative cooling is important (as 
in the case of galaxy clusters discussed in \secref{sec:cores}), 
the electron temperature equation should also have a cooling 
term, $-n_in_e\Lambda(T_e)$, where $\Lambda$ is the cooling function 
\citep[e.g.,][]{Tozzi}. 

Note that, in principle, since we kept the FLR terms in the pressure 
tensor, we should also have kept FLR corrections in the CGL equations. These arise 
from the FLR contribution to the heat flux --- in the collisional limit, 
it is the usual diamagnetic heat flux 
$\delta\vq = (5n\vth^2/4\Omega)\,\vb\times\vdel T$ \citep[see][]{Braginskii}. 
While the unperturbed part of these FLR terms is small compared 
to other macroscale terms, their perturbed part  
is comparable to the perturbed gyroviscous stress terms ($\vk\cdot\delta\vG$ 
in \eqref{eq:du}). In \eqref{eq:du}, the diamagnetic heat-flux terms 
are part of the perturbed pressures $\dpperp$ and $\dppar$. Since 
the instabilities we study in this paper are Alfv\'enically polarized 
and so are indifferent to pressure perturbations, we do not need 
to calculate the diamagnetic heat fluxes and, therefore, omit them.  

\section{Flux freezing}
\label{app:flux}

The non-MHD terms in \eqref{eq:ind} will still preserve 
the magnetic-field topology if the electric field can be 
expressed in the form $\vE = -\vueff\times\vB/c + \vdel\chi$, 
where $\chi$ is an arbitrary scalar function and 
$\vueff$ is some effective velocity field into which the flux 
will be frozen. Consider \eqref{eq:fbal}. It is not hard 
to show that the electron pressure term is
\bea
\nonumber
\lefteqn{\vdel\cdot\vP_e = \vdel\pperpe - \vdel\cdot\lt[\vb\vb(\pperpe-\ppare)\rt]}&&\\
\nonumber
&&\lefteqn{= \vdperp(\pperpe-\ppare) - (\pperpe-\ppare)\lt(\frac{\vdperp B}{B} + \vb\cdot\vdel\vb\rt)}\\
&&\lefteqn{+\vdel\ppare + (\pperpe-\ppare)\,\frac{\vdel B}{B}},
\eea
where $\vdperp=(\vI-\vb\vb)\cdot\vdel$ and we have used 
$\vdel\cdot\vb=-\vb\cdot\vdel B/B$. 
Since the first two terms are perpendicular 
to $\vB$, they can be represented as a vector product of some effective 
vector field with $\vB$. Therefore, introducing 
\bea
\nonumber
\vueff &=& \vu_e-\frac{c}{e n_eB^2}\Bigl[\vdperp(\pperpe-\ppare)\Bigr.\\ 
&&-\Bigl.(\pperpe-\ppare)\lt(\frac{\vdperp B}{B} + \vb\cdot\vdel\vb\rt)\Bigr]\times\vB,
\eea
where $\vu_e = \vu - (c/4\pi en_e)\vdel\times\vB$, we find 
from \eqref{eq:fbal}
\beq
\vE = -\frac{\vueff\times\vB}{c} - \frac{\vdel\ppare}{en_e}
- \frac{\pperpe-\ppare}{en_e}\frac{\vdel B}{B}. 
\eeq
With this electric field, Faraday's law \exref{eq:Faraday} becomes
\bea
\nonumber
\frac{\dd\vB}{\dd t} &=& \vdel\times(\vueff\times\vB)
- \frac{c\vdel n_e\times\vdel\ppare}{en_e^2}\\
&& + \lt(c\vdel\frac{\pperpe-\ppare}{en_e}\rt)\times\frac{\vdel B}{B},
\label{eq:Bueff}
\eea
which is an equivalent form of \eqref{eq:ind}. 
Thus, the magnetic field is frozen into the effective velocity 
field $\vueff$ except for two effects. 
The first of the two non-flux-conserving terms in \eqref{eq:Bueff}
is the well-known \citet{Biermann} battery (believed to be one of the 
mechanisms responsible for seeding the cosmic plasma with the initial 
magnetic fluctuations, subsequently amplified by turbulent dynamo; 
see \citealt{Kulsrud_BB}); the second is an effect due to the 
electron pressure anisotropy. While it is not a battery 
in the sense of producing magnetic field from a zero initial 
condition (plasma is unmagnetized when $B=0$, so there is 
no pressure anisotropy), it is independent of the field strength
and, therefore, can act as a source term. 

The flux unfreezing effect due to the electron pressure 
anisotropy is small except at very small scales. 
If we use \eqref{eq:aniso} to estimate 
$\pperpe-\ppare\sim (n_eT_e/\nu_{ei})(1/B)\did B/\did t$, 
we find, very roughly, that the flux conservation is 
significantly violated only if the scale of variation 
of $T_e$ and $B$ {\em perpendicular} to $\vB$ is 
$l_\perp\lesssim (\rho_e\mfp)^{1/2}$. While this is larger 
than the electron inertial or Larmor scales, where flux 
normally unfreezes in a collisionless plasma, and can be 
larger than the Ohmic resistive scale, it is still extremely 
small compared to any scales relevant for the macroscopic 
dynamics (for the reference cluster 
core parameters used in \secref{sec:cores}, 
we get $l_\perp\sim 10^9$~km). 
Note that the parallel firehose and gyrothermal instabilities 
considered in the main part of this paper are unaffected 
by this flux unfreezing effect because we set $\kperp=0$ 
and the instabilities contained no perturbation of the 
magnetic-field strength. 

\label{lastpage}

\end{document}